\documentclass[11pt]{article}

\usepackage[margin=1in]{geometry}
\usepackage{amsmath}
\usepackage{booktabs}
\usepackage{graphicx}
\usepackage{xurl}
\usepackage{hyperref}
\usepackage{enumitem}
\usepackage{xcolor}
\usepackage{array}
\usepackage{tikz}
\usepackage{listings}
\usepackage{placeins}
\usetikzlibrary{arrows.meta,calc,positioning}

\title{SWE-Future: Forecast-Conditioned Data Synthesis for Future-Oriented Software Engineering Agents}
\author{Qiao Zhao \quad JianYing Qu \quad Jun Zhang \quad Yehua Yang \quad Hanwen Du \quad Zhongkai Sun\\Baidu Inc.}
\date{June 9, 2026}

\newcommand{\swefuture}{\textsc{SWE-Future}}
\newcolumntype{L}[1]{>{\raggedright\arraybackslash}p{#1}}
\begin{document}
\maketitle

\begin{abstract}
Realistic coding-agent benchmarks often replay public GitHub issues and pull
requests, making them vulnerable to overlap with model pretraining,
fine-tuning, synthetic-data generation, or benchmark-driven model selection.
Fully synthetic tasks avoid direct historical replay, but can drift away from
real repository needs. We propose \swefuture{}, a forecast-conditioned data
synthesis method for future-oriented coding tasks. Given a forecast snapshot at
time $T_0$, the method uses only pre-$T_0$ repository evidence to forecast future
feature implementation/enhancement, bugfix, and refactor task families. We first
validate this forecasting step retrospectively: after forecasts are fixed, later
pull requests are used only to measure whether the predicted task families match
future repository work. In an 80-repository study, the forecaster achieves
58.1\% future-work relevance under the main semantic matching metric. We then use
validated forecast families as conditioning signals to synthesize a 200-task
coding-agent dataset across 61 repositories from a task-generation snapshot,
rather than replaying the later pull requests used for validation. \swefuture{}
shows that repository-evolution forecasts can guide realistic, future-oriented
coding-task synthesis while reducing direct dependence on historical
pull-request replay.
\end{abstract}

\section{Introduction}

Software engineering agents need training and evaluation tasks that are
realistic, repository-specific, and aligned with how active projects evolve.
Many strong coding-agent datasets and benchmarks get this realism by replaying
public GitHub issues and pull requests~\cite{jimenez2024swebench,
wang2025swebenchpp,li2025feabench,guo2026swefactory}.
That strategy is valuable, but it creates a growing validity problem: public
repositories, benchmark releases, and evaluation traces may appear in
pretraining, fine-tuning, synthetic-data generation, or model-selection loops.
For widely tracked coding-agent benchmarks, direct historical replay therefore
carries data-contamination risk for rapidly evolving coding
models~\cite{riddell2024contamination,matton2024leakage,badertdinov2025swerebench}.
Pure synthetic generation can reduce direct replay, but it often loses
repository pressure. A synthetic task may be plausible in isolation while
ignoring project conventions, maintainer priorities, realistic dependency
constraints, or the kinds of tests that the repository would actually accept.

\swefuture{} targets the middle ground. Instead of replaying historical pull
requests or asking an agent to invent arbitrary tasks, it first predicts likely
future repository evolution from pre-snapshot evidence. The validated prediction
unit is a task family, not an exact PR title. A family can describe, for example,
``Zarr/backend capability growth in xarray'', ``false-positive fixes in
pylint'', or ``callback API extension in Dash''. The final synthesis step then
uses a task-generation snapshot, denoted $T_{\mathrm{gen}}$, and the validated
family as a conditioning signal to construct future-oriented coding tasks.

\begin{figure}[!h]
\centering
\includegraphics[width=\linewidth]{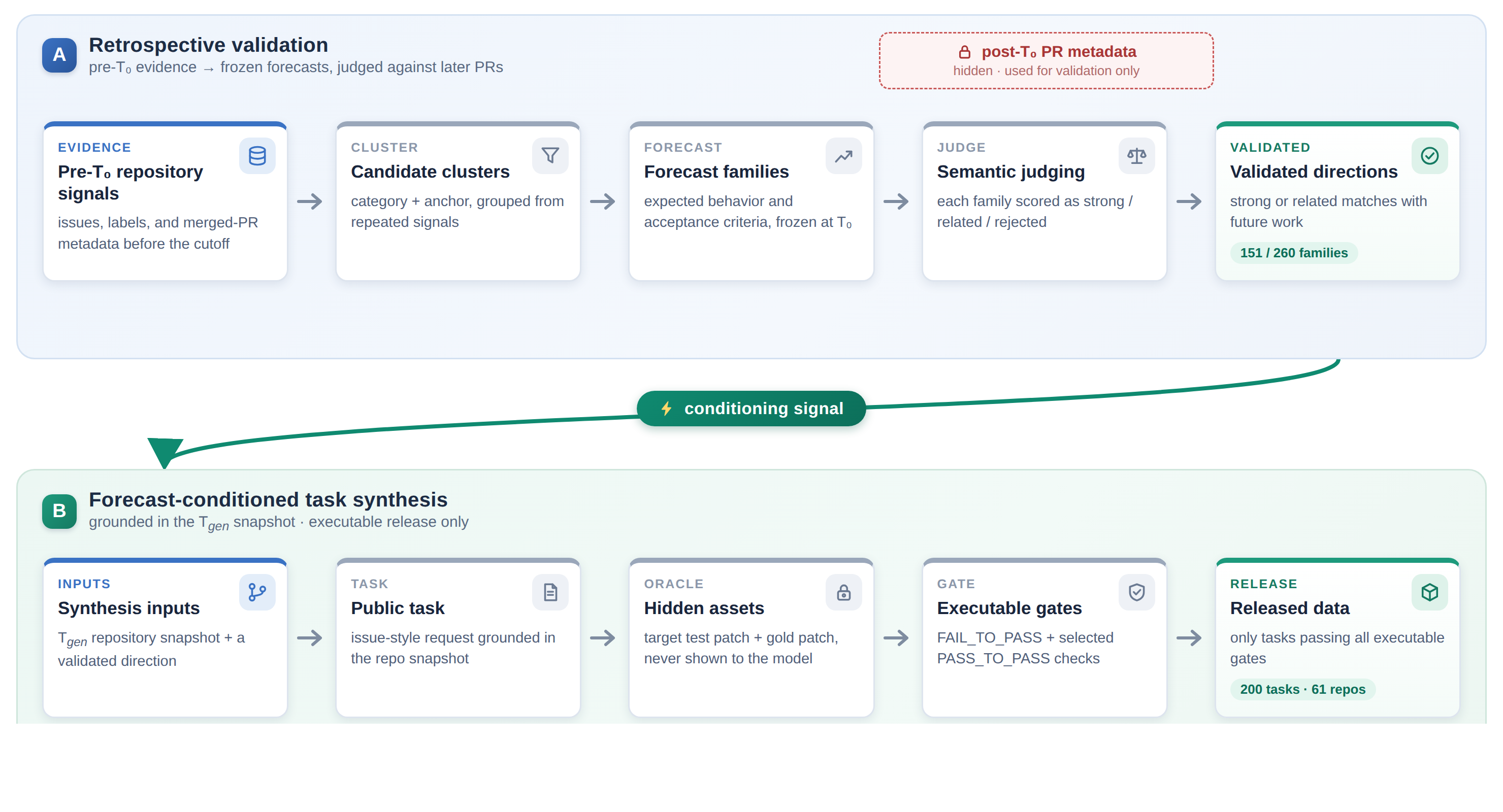}
\caption{\swefuture{} overview. The top lane validates frozen forecast
families: pre-$T_0$ repository evidence produces task-family forecasts, and
post-$T_0$ PR metadata is used only to judge semantic matches. The bottom lane
synthesizes tasks from the $T_{\mathrm{gen}}$ repository snapshot and validated
directions, then releases only tasks with target-test and gold-patch executable
evidence.}
\label{fig:pipeline}
\end{figure}

The method has four stages. First, for each repository snapshot at time $T_0$,
\swefuture{} builds an evidence bundle from pre-$T_0$ issues, pull requests,
labels, and text. Second, it forecasts feature implementation/enhancement,
bugfix, and refactor task families from that evidence. Third, in a retrospective
study only, it freezes the forecasts and compares them with later pull-request
metadata to measure whether the predicted families match future repository work.
Fourth, at $T_{\mathrm{gen}}$, it uses validated families as conditioning
signals to synthesize coding tasks from the repository state visible to the task
constructor, without directly replaying historical pull requests.
Figure~\ref{fig:pipeline} in the method section summarizes this information
boundary.

This paper makes three contributions. First, it introduces
forecast-conditioned data synthesis as a construction method for realistic,
future-oriented coding-agent tasks. Second, it gives a retrospective validation
procedure for testing whether forecast families capture future repository work;
in an 80-repository study, the forecaster reaches 58.1\% future-work relevance
under the main semantic matching metric. Third, it demonstrates forecast-to-task
conversion by grounding validated families in $T_{\mathrm{gen}}$ repository
snapshots and producing a 200-task coding-agent dataset across 61 repositories
with executable validation.

\section{Related Work}

\paragraph{Historical GitHub task construction.}
SWE-Bench established repository-level issue resolution as a standard coding
agent evaluation setting by drawing tasks from real GitHub issues and their
corresponding pull requests~\cite{jimenez2024swebench}. SWE-Bench++ scales this
family of GitHub-derived evaluation across more repositories and languages, and
FEA-Bench narrows the target to feature implementation by collecting
feature-focused pull requests and relevant tests~\cite{wang2025swebenchpp,li2025feabench}.
SWE-Factory further automates GitHub issue-resolution data construction through
test recovery, environment building, and fail-to-pass validation~\cite{guo2026swefactory}.
These systems provide realism because the target work has already happened.
\swefuture{} instead uses post-$T_0$ pull requests only as hidden retrospective
evidence for frozen forecasts; public tasks are generated from a
task-generation repository snapshot and a validated forecast family.

\paragraph{Benchmark contamination and leaderboard pressure.}
Public coding benchmarks are useful precisely because they create shared targets
for model developers. That same visibility creates contamination risk: benchmark
repositories, issue text, pull-request metadata, reference solutions, or
derivative evaluation traces can enter pretraining, fine-tuning, synthetic-data,
or model-selection pipelines~\cite{riddell2024contamination,matton2024leakage,
badertdinov2025swerebench}.
This risk is especially salient for widely used GitHub-derived benchmarks such
as SWE-Bench: model developers can track public leaderboards, optimize prompts
or data against benchmark behavior, and repeatedly inspect public source
projects and task texts. Our goal is not to prove that model parameters have
never seen a repository; it is to avoid direct replay of observed historical
pull requests as task material.

\paragraph{Task factories and synthetic environments.}
SWE-Hub and daVinci-Env focus on large-scale environment construction and
executable task generation, while SWE-Playground synthesizes projects, tasks,
tests, and implementations from scratch to improve scale and
control~\cite{zeng2026swehub,fu2026davincienv,zhu2026sweplayground}. These
systems reduce the engineering cost of runnable tasks; \swefuture{} targets the
upstream question of which repository-specific task families should be
synthesized before task construction, using forecast validity as the
conditioning signal.

\paragraph{Software evolution and temporal validity.}
Software engineering has long used repository histories to predict risky modules,
defect-prone changes, and likely co-changes~\cite{graves2000faults,
nagappan2005churn,hassan2009changecomplexity,kamei2013jit,
zimmermann2004versionhistories}.
Recent time-consistent benchmark work also emphasizes temporal discipline by
snapshotting a repository at $T_0$, constructing repository knowledge from
pre-$T_0$ artifacts, and evaluating on tasks derived from later pull
requests~\cite{sun2026timeconsistent}. \swefuture{} borrows this temporal
boundary but changes the target: it forecasts coarse future work families and
uses those families to guide synthesis at $T_{\mathrm{gen}}$.

Together, these lines of work leave a gap. Historical GitHub replay supplies
realism but increases direct contamination pressure; fully synthetic task
factories improve scale and control but can lose repository-specific demand; and
time-consistent evaluation still often relies on later realized work as the task
itself. \swefuture{} addresses this gap by using later work only to validate
forecast families, then synthesizing tasks from a task-generation snapshot
conditioned on those families.

\section{Method}

\begin{figure}[!h]
\centering
\includegraphics[width=\linewidth]{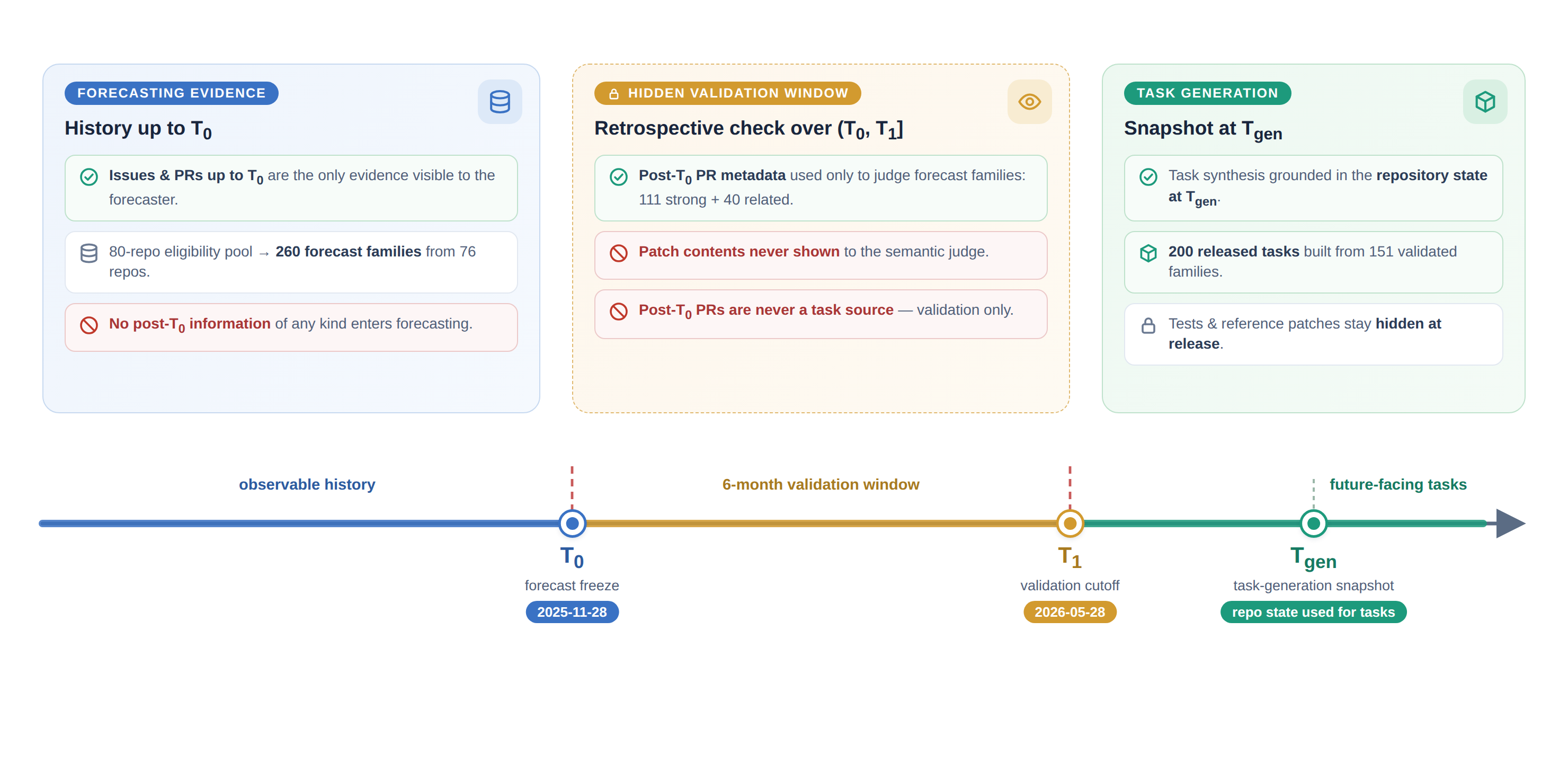}
\caption{Temporal information boundary in \swefuture{}. Forecasting uses only
repository evidence visible at or before $T_0$; retrospective validation uses
post-$T_0$ PR metadata only to judge frozen forecast families; task synthesis is
grounded in the $T_{\mathrm{gen}}$ repository snapshot, while task-specific
tests and gold patches remain hidden evaluation assets.}
\label{fig:timeline}
\end{figure}

\swefuture{} is a forecast-conditioned data synthesis method for constructing
repository-specific coding-agent tasks without directly replaying realized pull
requests. For each repository $R$, the method separates three time points: a
forecast snapshot $T_0$, a retrospective validation horizon $(T_0,T_1]$, and a
task-generation snapshot $T_{\mathrm{gen}}$. Evidence at or before $T_0$ is used
to predict future task families; post-$T_0$ PR metadata is used only to validate
whether those frozen forecasts correspond to later repository work; and task
generation is performed from the repository state visible at
$T_{\mathrm{gen}}$.
Figure~\ref{fig:timeline} makes this temporal boundary explicit.

\FloatBarrier
\FloatBarrier

The method has four stages. It first builds task-family forecasts from
pre-$T_0$ repository evidence. It then validates the frozen families against
post-$T_0$ PR metadata using semantic matching. Next, it uses validated families
as conditioning signals to synthesize issue-style public task specifications
from the $T_{\mathrm{gen}}$ repository snapshot. Finally, it admits only tasks
that pass target-test construction, leakage checks, \texttt{FAIL\_TO\_PASS}, and
selected \texttt{PASS\_TO\_PASS} validation.

The forecast unit is a family rather than a realized pull request:
\[
  f_i = (\text{category}, \text{family anchor}, \text{expected behavior},
  \text{evidence refs}, \text{target hints}, \text{acceptance criteria}).
\]
A family describes a repository demand direction, such as a recurring bug class,
an expected feature implementation/enhancement, or a refactoring need. In this
paper we keep feature implementation/enhancement, bugfix, and refactor families,
and exclude dependency/platform updates, CI/release work, documentation-only
changes, formatting-only edits, and test-only cleanup. This scope keeps the
method focused on coding-agent data synthesis rather than general maintenance
activity prediction.

\subsection{Forecasting and Retrospective Validation}

\paragraph{Repository pool.}
The retrospective validation pool contains 80 repositories. We begin from Python
repositories covered by or adjacent to SWE-bench and the SWE-bench harness, then
expand to active Python and software-engineering ecosystem projects across
testing, linting, packaging, web and networking, data and scientific computing,
AI and agent frameworks, and infrastructure. Candidate repositories are first
filtered by merged-PR activity in the common retrospective window, because
forecast validation requires enough post-$T_0$ repository work to compare against
frozen forecasts. We then require local GitHub history payloads containing
pre-$T_0$ issue and PR signals for forecasting and post-$T_0$ PR metadata for
retrospective validation.

The remaining candidates are ranked by the density of post-$T_0$ work in the
retained categories: feature implementation/enhancement, bugfix, and refactor,
together with the amount of usable pre-$T_0$ evidence. Dependency/platform,
CI/release, documentation-only, formatting-only, and test-only activity is not
treated as positive evidence unless it also contains direct bug or regression
evidence.
Repositories with very few post-$T_0$ PRs or very few in-scope post-$T_0$ PRs
are down-weighted. The final pool is selected to provide enough post-$T_0$ work
for retrospective validation and to maintain domain diversity. It is not meant
to be an unbiased sample of GitHub repositories. All repositories use the same
retrospective window:
\[
T_0=\text{2025-11-28}, \qquad T_1=\text{2026-05-28}.
\]

\paragraph{Forecast construction.}
Forecast construction uses four explicit steps, all restricted to information
visible at or before $T_0$. The forecasting assumption is persistence-based:
repeated pre-$T_0$ demand around the same repository-specific anchor is treated
as evidence of a continuing work direction. The method extrapolates that
direction into a future-oriented family specification rather than a concrete PR,
patch, or task.

\textbf{Step 1: Collect pre-$T_0$ signals.}
For each repository, the forecaster reads issue and merged-PR records visible
before $T_0$, using labels, titles, and short body excerpts as evidence. It
applies rule-based exclusion filters to remove dependency bumps, platform
updates, CI/release work, documentation-only edits, formatting-only edits, and
test-only cleanup. If a record contains a maintenance keyword but also direct bug
evidence, such as regression, crash, incorrect behavior, or exception, it is
kept for the bugfix channel. The remaining records are assigned to feature
implementation/enhancement, bugfix, or refactor using category keyword patterns
over the title, labels, and body excerpt.

\textbf{Step 2: Cluster signals into candidate directions.}
The forecaster extracts project-specific anchor terms from labels, titles, module
mentions, and short body excerpts. Generic project-management words, repository
boilerplate, URLs, overly long path fragments, and noisy anchors are removed.
Each retained record contributes its strongest anchors, and the first three
anchors are used to form clusters keyed by $(\text{category}, \text{anchor})$.
A candidate cluster is considered only when it contains at least two pre-$T_0$
signals. This requirement prevents an isolated mention from becoming a forecast
family by itself.

\textbf{Step 3: Score and filter clusters.}
Candidate clusters are ranked with a fixed heuristic score:
\[
s = n + 0.45 n_{\mathrm{issue}} + 0.25 n_{\mathrm{prePR}} +
0.05 n_{\mathrm{label}},
\]
where $n$ is the number of clustered signals, $n_{\mathrm{issue}}$ is the
number of issue signals, $n_{\mathrm{prePR}}$ is the number of pre-$T_0$ merged-PR
signals, and $n_{\mathrm{label}}$ is the number of distinct labels in the
cluster. These weights are not learned parameters; they are a deterministic
ranking rule used in this study.

\begin{center}
\small
\begin{tabular}{@{}L{0.24\linewidth}L{0.16\linewidth}L{0.50\linewidth}@{}}
\toprule
Term & Weight & Rationale \\
\midrule
$n$ & 1.0 & Base evidence volume from repeated signals. \\
$n_{\mathrm{issue}}$ & +0.45 & Issues state user-visible demand. \\
$n_{\mathrm{prePR}}$ & +0.25 & Pre-$T_0$ merged PRs show project precedent. \\
$n_{\mathrm{label}}$ & +0.05 & Label diversity is used only as a weak tie breaker. \\
\bottomrule
\end{tabular}
\end{center}

Clusters below 2.8 are discarded. The remaining clusters are sorted by score,
capped at five families per repository, and capped at two families per category.
The confidence shown for a family is a deterministic display score derived from
the cluster score, $\min(0.91, 0.54 + 0.045s)$, rather than a learned
probability.

\textbf{Step 4: Write each cluster as a forecast family.}
For each selected cluster, the forecaster instantiates a category-specific
template using the cluster anchor: feature implementation/enhancement clusters
become capability, API, provider, or format-extension families; bugfix clusters
become recurring failure, edge-case, regression, or incorrect-behavior families;
and refactor clusters become behavior-preserving restructuring families. The
emitted family contains the category, family anchor, expected behavior,
acceptance checklist, target hints, supporting pre-$T_0$ signal references, and
non-goals.
Appendix~\ref{app:forecast-templates} lists the templates used for this
cluster-to-family step, and Appendix~\ref{app:forecast-examples} gives concrete
forecast examples.

The forecaster emitted 260 families across 76 of 80 repositories; the remaining
four repositories abstained because no cluster crossed the evidence threshold.
Each emitted family is a structured forecast specification rather than a task,
and it is the only forecast-side input available to later task synthesis.

\begin{table}[t]
\centering
\small
\begin{tabular}{@{}lrr@{}}
\toprule
Category & Forecast families & Share \\
\midrule
Bugfix & 139 & 53.5\% \\
Feature impl./enh. & 93 & 35.8\% \\
Refactor & 28 & 10.8\% \\
\midrule
All & 260 & 100.0\% \\
\bottomrule
\end{tabular}
\caption{Feature implementation/enhancement, bugfix, and refactor family
forecast yield on the 80-repository eligibility pool.}
\end{table}

\paragraph{Retrospective validation.}
After forecasts are frozen, retrospective validation checks whether they match
later repository work in the six-month window $(T_0,T_1]$, ending at
$T_1=\text{2026-05-28}$. The endpoint $T_1$ closes the validation window; it is
not a point after which new validation evidence is gathered. Validation has two
stages.

\textbf{Retrieval.}
For each forecast family, a matcher retrieves post-$T_0$ PR candidates from the
same repository. It applies the same retained-category scope filter: dependency,
platform, CI/release, documentation-only, formatting-only, and test-only PRs are
removed unless their text also contains direct bug or regression evidence. A
post-$T_0$ PR remains a retrieval candidate only if it shares the forecast
category or contains the forecast anchor. Candidate pairs are ranked by
\[
m(f,p)=0.45J(f,p)+0.12S_{\mathrm{title}}(f,p)+B_{\mathrm{cat}}(f,p)+B_{\mathrm{anchor}}(f,p),
\]
where $J$ is token Jaccard overlap between the forecast-family text and PR
metadata, $S_{\mathrm{title}}$ is title similarity, $B_{\mathrm{cat}}$ is 0.18
for category agreement and 0.04 for a classifiable category mismatch, and
$B_{\mathrm{anchor}}=\min(0.34,0.12h)$ for $h$ anchor or target-module hits in
the PR text. Retrieval forms a top-3 shortlist; semantic validation uses the
top-1 candidate. Post-$T_0$ patches are never shown to the judge.

\textbf{Judging.}
The primary semantic judge labels each frozen forecast and top-1 post-$T_0$ PR
metadata on a single relevance dimension with three ordered tiers: strong,
related, and rejected. Strong means the forecast directly matches later work,
including exact, partial, and event-family hits. Related means the forecast is
not the same later task but still matches the category or module direction.
Rejected covers no-match and unclear cases. The main metric is forecast-family
relevance:
\[
  \frac{\#\text{strong} + \#\text{related}}{\#\text{forecast families}},
\]
which measures synthesis eligibility rather than exact post-$T_0$ PR prediction.
This metric counts related families because the downstream goal is not to
predict the exact post-$T_0$ PR, but to identify repository-natural directions that
can condition task synthesis. A related family is therefore useful when it points
to the same category or module pressure even if it does not match a single later
PR exactly.

The resulting forecast-family relevance is 151/260 (58.1\%). The stricter
strong hit rate is 111/260 (42.7\%). There are 61 repositories with at least one
strong+related family. Bugfix families are currently the strongest source of
eligible synthesis directions, with 89/139 strong+related; feature
implementation/enhancement families are harder, with 45/93 strong+related.

An independent semantic audit over all 260 forecasts agrees with the primary
relevance-vs-rejected decision in 216/260 cases (83.1\%). We use this agreement
as a consistency check, while keeping 151/260 as the main future-work relevance
metric.

\begin{figure}[!h]
\centering
\includegraphics[width=\linewidth]{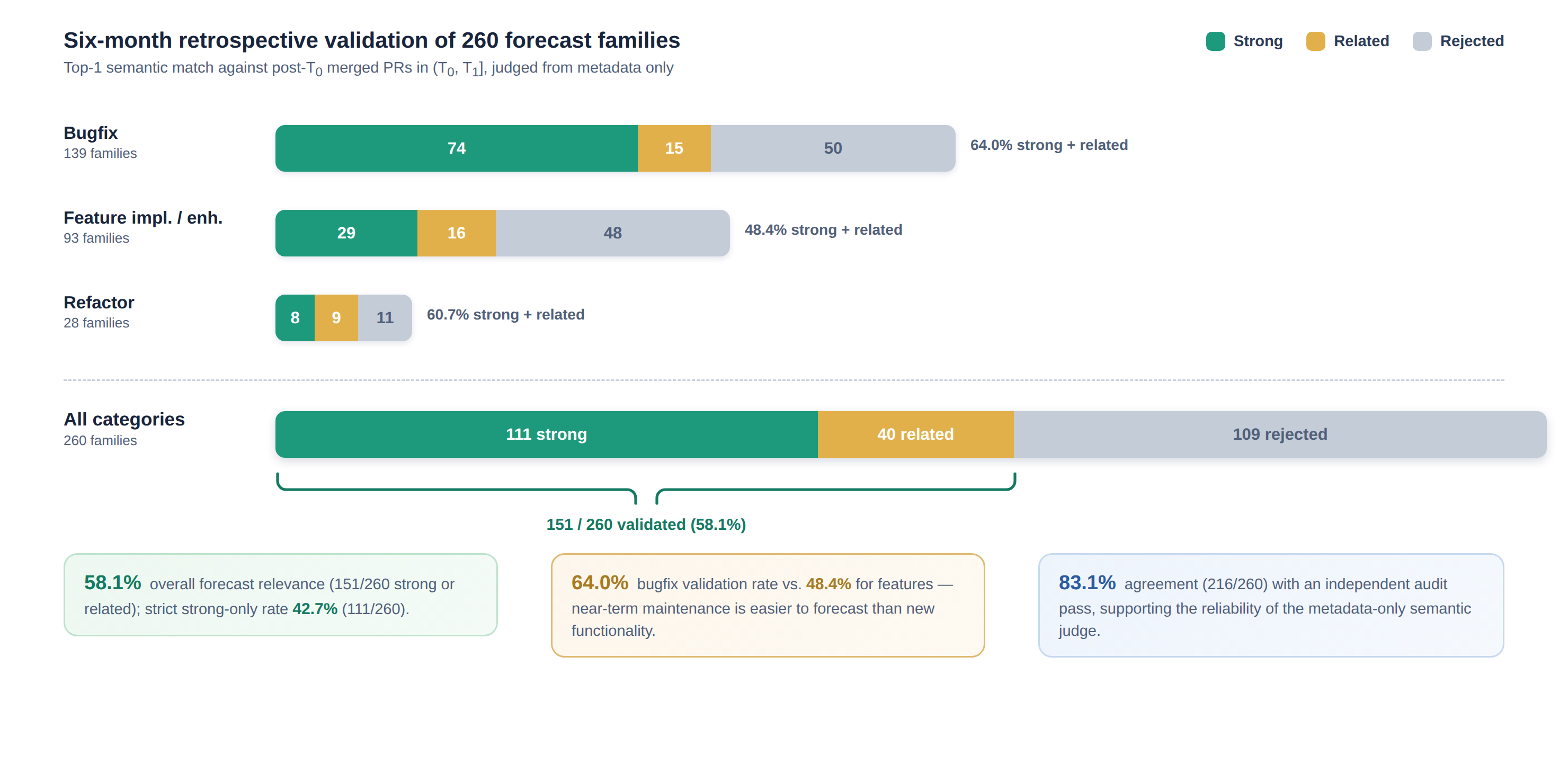}
\caption{Retrospective validation outcome for 260 forecast families. Bars show
strong, related, and rejected top-1 semantic matches against post-$T_0$ merged PR
metadata in $(T_0,T_1]$; the main relevance metric counts strong+related
families.}
\label{fig:validation}
\end{figure}

\FloatBarrier

\subsection{Task Generation and Data Construction}

Forecast families that pass the retrospective relevance check, either as strong
or related matches, become conditioning signals for task synthesis at
$T_{\mathrm{gen}}$. The forecast family supplies the repository-specific
direction; inspection of the $T_{\mathrm{gen}}$ codebase turns that direction
into a concrete change request.

This separation is central to the construction. Post-$T_0$ pull requests are
used only at the forecast-family validation stage, after forecasts are frozen,
to test whether a predicted direction matches later repository work. They are a
family-level validation signal, not a task source. The 200 released tasks do not
have target PRs: each task is synthesized from the $T_{\mathrm{gen}}$ repository
snapshot and repository evidence visible to the constructor. The generator
therefore uses retrospective validation only to select reliable directions, not
to replay or transform post-$T_0$ PR artifacts into prompts, tests, or gold
patches.

\subsubsection{From Forecast Families to Tasks}

The synthesis stage materializes a 200-task coding-agent dataset across 61
repositories. The 151 strong-or-related forecast families do not map one-to-one
onto tasks. A family can contribute multiple task rows when
$T_{\mathrm{gen}}$ code inspection exposes separable behavior gaps, capability
extensions, or refactor targets; it can also contribute no released row if no
executable oracle survives construction. The 200-task mix is used as a
candidate-construction target, not a release shortcut: the queue is built with
caps per repository and per forecast family, with the related tier capped so
that strong matches dominate the pool. Rows that fail target audit, leakage
review, or executable validation are revised or replaced rather than admitted to
satisfy a quota. We allocate 120 bugfix, 60 feature implementation/enhancement,
and 20 refactor tasks because bugfix and feature directions more often yield
behavior-level oracles, while refactors are harder to validate with focused
executable checks.

\begin{figure}[!h]
\centering
\includegraphics[width=\linewidth,trim=0 120bp 0 260bp,clip]{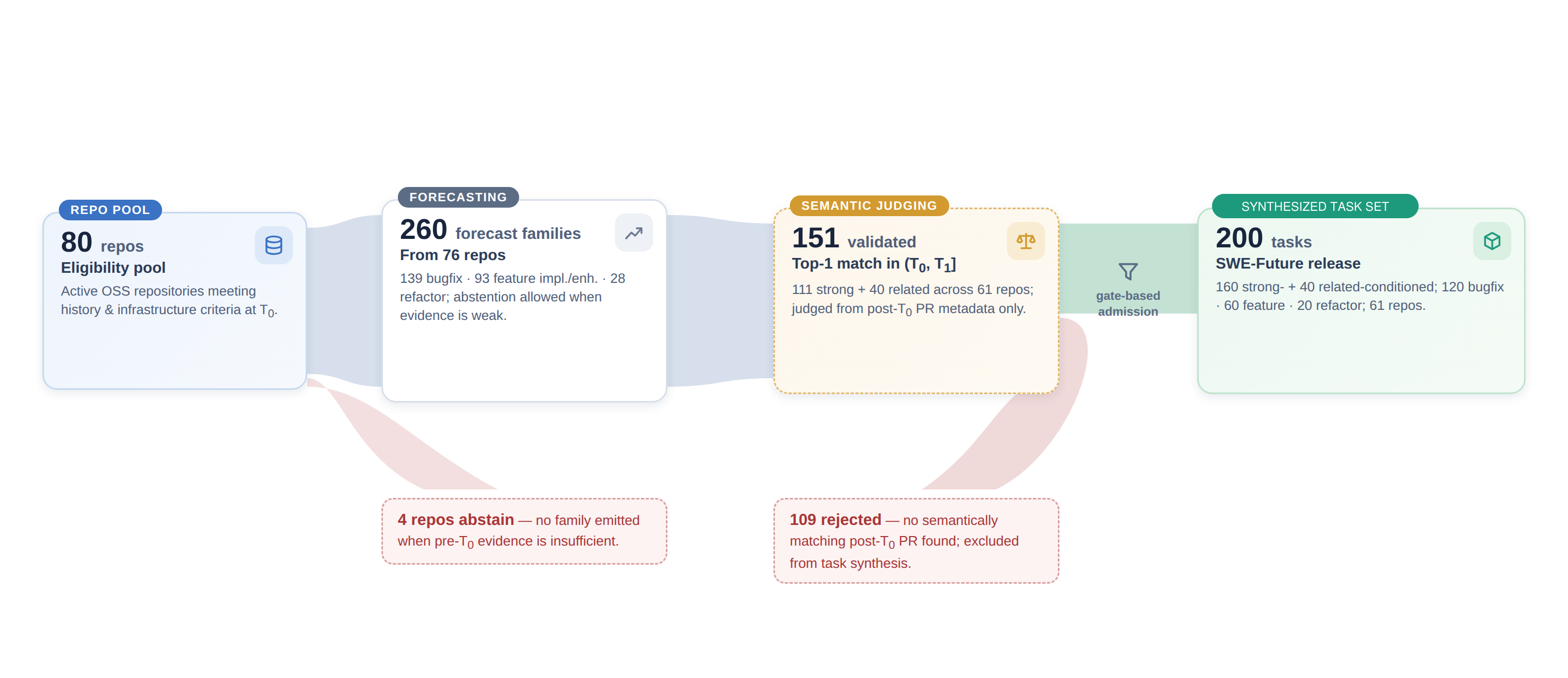}
\caption{Forecast-to-task selection funnel. The 80-repository pool yields 260
forecast families from 76 repositories; 151 families validate as strong or
related under semantic judging and condition the 200 synthesized task set through
gate-based admission.}
\label{fig:funnel}
\end{figure}

Task synthesis follows the same three-stage route for every row:
\begin{enumerate}[leftmargin=*]
\item \textbf{Select a validated direction.} Choose a strong or related forecast
family, keep its pre-$T_0$ evidence references as conditioning context, and load
the corresponding task-generation snapshot, meaning the repository state visible
to the constructor rather than validation artifacts outside that snapshot.
\item \textbf{Ground it in the $T_{\mathrm{gen}}$ codebase.} Inspect code, public
APIs, tests, and project conventions to identify a concrete behavior gap,
capability extension, or behavior-preserving refactor that is natural for that
project.
\item \textbf{Release only executable tasks.} Write a normal issue-style request,
construct the test patch and gold patch, and admit the task only after
leakage review, \texttt{FAIL\_TO\_PASS} validation, and selected preservation
checks.
\end{enumerate}

The grounding and executable-release stages are organized as a multi-agent
workflow, with separate artifacts for context selection, task writing, oracle
design, patch construction, and verification. The loop starts from a validated
family and searches the $T_{\mathrm{gen}}$ repository tree for
implementation and test candidates using the family anchor, target-module hints,
task category, and variant tokens. If no plausible implementation or test target
is found, the row is sent back for repository inspection rather than patch
generation.

\begin{figure}[t]
\centering
\includegraphics[width=\linewidth]{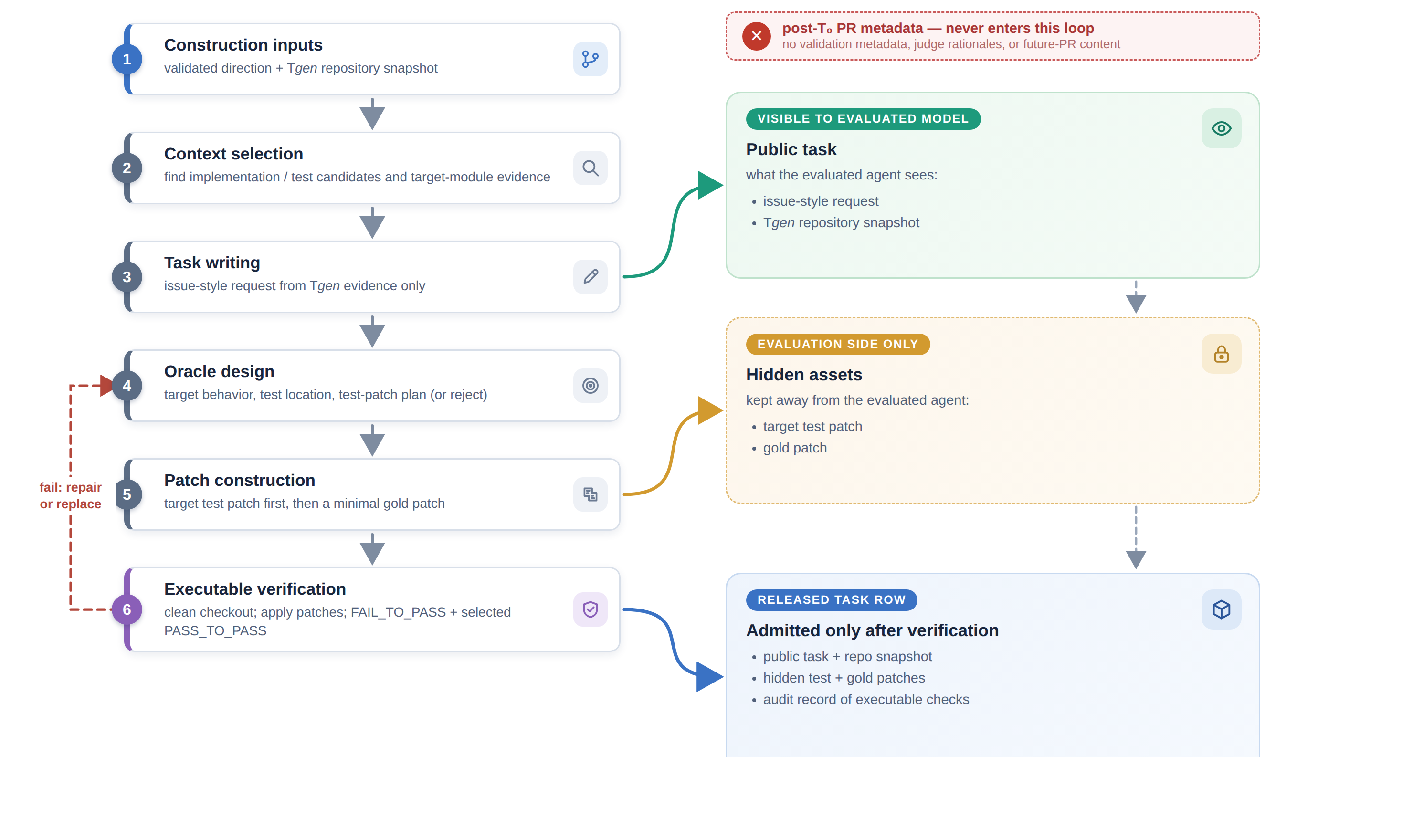}
\caption{Multi-agent task construction. Validated directions and the
$T_{\mathrm{gen}}$ snapshot enter a construction loop with separate context
selection, task-writing, oracle-design, patch-construction, and verification
roles. Only the public task and repository snapshot are model-visible; target
tests, gold patches, and verification logs remain hidden evaluation assets.}
\label{fig:construction-workflow}
\end{figure}

Next, a task-writing role produces the model-visible issue-style request from
the forecast family and $T_{\mathrm{gen}}$ evidence only. This role does not see
post-$T_0$ PR metadata, judge rationales, test-patch drafts, or gold-patch
content. An oracle-design role then converts the request into a concrete target
behavior, a test location, and a test-patch plan; it can reject tasks that are
too broad, already solved in the snapshot, or not testable with a focused
oracle. A patch-construction role writes the test patch first and then a gold
patch scoped to the smallest implementation path. Finally, a verification role
starts from a clean checkout, checks that both patches apply, runs the
\texttt{FAIL\_TO\_PASS} sequence, runs selected preservation checks, and either
admits the row or returns it to the earlier roles for repair. Table
\ref{tab:construction-roles} summarizes the artifacts, and
Figure~\ref{fig:construction-workflow} summarizes the construction workflow and
its public/hidden artifact boundary. Appendix~\ref{app:task-construction-details}
details the role strategies, and Appendix~\ref{app:synthesized-task-example}
shows a complete synthesized task row.

This separation matters because the task writer should not know the realized
post-$T_0$ fix or the hidden oracle. Without that barrier, the public task could
be phrased around a known solution and become a disguised historical replay.
Task-specific tests, gold patches, provenance records, validation labels, and
execution logs are therefore kept as hidden construction and evaluation assets.

\begin{table}[t]
\centering
\small
\begin{tabular}{@{}L{0.19\linewidth}L{0.34\linewidth}L{0.38\linewidth}@{}}
\toprule
Role & Input & Output or gate \\
\midrule
Context selection & Forecast family and task-generation snapshot &
Implementation/test candidates and target-module evidence \\
Task writing & Forecast family plus $T_{\mathrm{gen}}$ repository evidence &
Model-visible issue statement without validation metadata, judge rationales, or
test/gold patch material \\
Oracle design & Public task, target files, and $T_{\mathrm{gen}}$ code excerpts &
Behavior under test, test-patch plan, gold-patch scope, or rejection \\
Patch construction & Oracle plan and task-generation snapshot &
Task-specific test patch and gold patch \\
Verification & Public task, test patch, and gold patch &
\texttt{FAIL\_TO\_PASS} result plus selected \texttt{PASS\_TO\_PASS} checks \\
\bottomrule
\end{tabular}
\caption{Multi-agent construction workflow for grounding and executable release.}
\label{tab:construction-roles}
\end{table}

\begin{table}[t]
\centering
\small
\begin{tabular}{@{}L{0.31\linewidth}rL{0.41\linewidth}@{}}
\toprule
Dataset view & Tasks & Notes \\
\midrule
Bugfix & 120 & Correctness and edge-case repair tasks \\
Feature impl./enh. & 60 & Capability, API, format, and option-support tasks \\
Refactor & 20 & Behavior-preserving internal improvement tasks \\
\midrule
Strong-family conditioned & 160 & Conditioned on retrospective strong matches \\
Related-family conditioned & 40 & Conditioned on retrospective related matches \\
\midrule
All tasks & 200 & $T_{\mathrm{gen}}$ task specifications with executable validation assets \\
\bottomrule
\end{tabular}
\caption{\swefuture{} dataset composition. Tasks are generated from
$T_{\mathrm{gen}}$ repository evidence and conditioned on validated forecast
families.}
\end{table}

\begin{table}[t]
\centering
\small
\begin{tabular}{@{}L{0.18\linewidth}L{0.18\linewidth}L{0.25\linewidth}L{0.27\linewidth}@{}}
\toprule
Repository & Category & Forecast family & Public task example \\
\midrule
\texttt{agronholm/anyio} & Bugfix & BlockingPortal edge-case failures &
Forward provider shutdown exceptions through a focused regression and minimal
BlockingPortal cleanup-path fix. \\
\texttt{celery/kombu} & Feature & Redis broker capability growth &
Honor the Redis global key-prefix option for expiration commands so broker keys
remain isolated across shared databases. \\
\texttt{scrapy/scrapy} & Refactor & Cleanup helper extraction &
Extract shutdown-sentinel logic from \texttt{\_parallel\_asyncio()} while
preserving behavior under regression tests. \\
\bottomrule
\end{tabular}
\caption{Examples of forecast-conditioned public task specifications. The
examples show the synthetic public task surface only; validation assets and
internal provenance are hidden.}
\label{tab:task-examples}
\end{table}

\subsubsection{Task Surface and Hidden Validation Assets}

The released task exposes the task-generation repository snapshot together with
a model-visible issue-style request. The request states the expected behavior,
acceptance criteria, and non-goals, while the repository snapshot supplies the
code context needed by the evaluated coding agent. The task does not claim to
reproduce a known pull request, and it does not expose the validation or
construction artifacts used to build the dataset row.

Those internal artifacts are kept on the hidden side: forecast-family
provenance, retrospective validation labels, oracle plans, task-specific test
patches, gold patches, and execution logs. They are used to audit why a task
direction was admitted and to verify that the task has an executable target.

\subsubsection{Executable Dataset Validation}

Dataset admission is gate-based rather than quota-based. A task must pass three
checks before release. First, a target audit verifies that the request can be
justified from the task-generation snapshot. This audit must identify a concrete
implementation target, a test target or test location, and a narrow behavior
that can be checked without requiring full-project redesign.

Second, a leakage review checks the public task surface against hidden
construction and validation artifacts. The public task may name current files,
APIs, symptoms, and acceptance criteria, but it must not contain test-patch
logic, gold-patch details, retrospective validation labels, judge
rationales, or retrospective validation metadata. The hidden provenance file
records the forecast family and validation evidence for audit only; it is never
shown to the evaluated coding agent.

Third, executable validation requires two independently constructed artifacts:
a task-specific test patch and a gold patch. The test patch defines the
task-specific checks and is withheld from evaluated coding agents during
evaluation runs. The gold patch is a maintainer-quality solution patch used to
prove that the task is solvable and that the test patch is not ill-posed. It is
also withheld from evaluated agents and is not treated as the only possible
solution. In \swefuture{}, both patches are constructed during data generation
rather than extracted from a historical PR. Both patches must apply cleanly. The
target oracle is then evaluated as a \texttt{FAIL\_TO\_PASS} target: after
applying only the test patch, the task must fail on the task-generation
snapshot; after also applying the gold patch, the same oracle must pass. For
behavior-preserving refactors, the target may be a focused structural or
characterization oracle, but the gold patch must still preserve the relevant
public behavior. We also run selected \texttt{PASS\_TO\_PASS} checks to guard
against patches that satisfy the new target while breaking existing behavior.
These preservation checks are chosen from repository-native public tests that
cover the touched module or nearby behavior, pass on the task-generation
snapshot, and avoid unstable external services or unusually heavy full-project
setup.

This validation design makes the public dataset useful as coding-agent data
rather than just as a list of plausible issue descriptions. The forecast family
decides the repository-specific direction; $T_{\mathrm{gen}}$ code inspection
grounds the task in the repository; executable validation ensures that the task
has a concrete behavioral target.

\section{Discussion}

\swefuture{} depends on the quality of repository-evolution forecasts, and this
forecasting step remains uncertain. The forecaster can pick up noisy anchors
from labels, tracker terms, or repository-specific process language rather than
clean code modules. The generated test patches make each task executable, but
they may be less precise than tests written by long-term maintainers with deeper
project context. This limitation is especially visible for refactor tasks, where
the intended change is internal and behavior-preserving rather than a new
externally observable behavior.

Future work should reduce uncertainty at both the forecast and execution levels.
On the forecasting side, stronger retrieval over repository evidence and better
noise filtering for labels and tracker language could improve the quality of
admitted forecast families. On the dataset-construction side, richer
project-specific test synthesis could make generated test patches closer to
maintainer-written tests.

\section{Conclusion}

\swefuture{} turns repository-evolution forecasting into a conditioning signal
for coding-agent data synthesis. The study shows that feature implementation,
bugfix, and refactor families can be forecast from pre-$T_0$ evidence with
measurable future relevance, and that these validated families can drive the
construction of 200 coding-agent tasks across 61 repositories. The central
contribution is a contamination-aware synthesis route: freeze repository
history, forecast likely project directions, ground those directions in
$T_{\mathrm{gen}}$ repository snapshots, and admit only tasks with executable
validation. Future improvements should focus on reducing forecast noise and
making generated test patches closer to maintainer-written tests.

\clearpage
\appendix

\section{Forecast Family Templates}
\label{app:forecast-templates}

The cluster-to-family step uses category-specific templates. These templates are
not task prompts; they convert a repeated pre-$T_0$ evidence cluster into a
coarse future-work direction that later task construction can ground in the
$T_{\mathrm{gen}}$ codebase.

\begin{center}
\small
\begin{tabular}{@{}L{0.24\linewidth}L{0.32\linewidth}L{0.34\linewidth}@{}}
\toprule
Category & Forecasted direction template & Acceptance emphasis \\
\midrule
Feature implementation/enhancement &
Extend user-visible \texttt{\{anchor\}} behavior, API coverage, provider
support, or format compatibility. &
Require a concrete capability or compatibility path, focused tests, and no
dependency/platform or documentation-only change. \\
Bugfix &
Fix recurring \texttt{\{anchor\}} failures, edge cases, regressions, or
incorrect behavior reported by users. &
Require a reproducible failure or edge case, a minimal implementation patch, and
a regression test that fails before and passes after the fix. \\
Refactor &
Restructure \texttt{\{anchor\}} internals, typing, deprecated paths, or helper
boundaries while preserving externally visible behavior. &
Require behavior preservation, local maintainability rationale, and tests or
checks that guard the preserved public behavior. \\
\bottomrule
\end{tabular}
\par\smallskip
\emph{Template summary used to turn high-scoring pre-$T_0$ clusters into
structured future-work directions.}
\end{center}

\clearpage
\section{Forecast Examples}
\label{app:forecast-examples}

The following examples show how repeated pre-$T_0$ signals become forecast
families and how those families are later judged against post-$T_0$ PR metadata.
These examples illustrate the method boundary: the post-$T_0$ PR is used only
for retrospective validation, not to construct the released public task.

\begin{center}
\footnotesize
\begin{tabular}{@{}L{0.19\linewidth}L{0.23\linewidth}L{0.24\linewidth}L{0.24\linewidth}@{}}
\toprule
Repository & Pre-$T_0$ cluster & Forecast family & Retrospective validation \\
\midrule
\texttt{PrefectHQ/prefect} &
Feature implementation/enhancement anchor \texttt{deployment}; source signals
include issues \#19037, \#18224, and \#19415. &
Extend \texttt{deployment} capability/API support. &
Strong: PR \#21954 added resizable columns on a deployments table, a
user-visible deployment enhancement. \\
\texttt{plotly/dash} &
Feature implementation/enhancement anchor \texttt{callback}; source signals
include pre-$T_0$ PRs \#3407, \#3397, and issue \#3493. &
Extend \texttt{callback} capability/API support. &
Strong: PR \#3563 added \texttt{hidden} as a configurable parameter for
clientside callbacks. \\
\texttt{celery/kombu} &
Feature implementation/enhancement anchor \texttt{broker}; source signals
include pre-$T_0$ PRs \#2408, \#2367, and issue \#2334. &
Extend broker capability/API support. &
Strong: PR \#2251 added Redis queue expiration support. \\
\texttt{Kludex/uvicorn} &
Bugfix anchor \texttt{hangs}; source signals include issues \#2735 and \#2754. &
Fix recurring \texttt{hangs} correctness and edge-case failures. &
Strong: PR \#2815 replaced fixed sleeps with polling in multiprocess tests and
\texttt{run\_server}, matching a hang/timing edge-case family. \\
\texttt{astropy/astropy} &
Bugfix anchor \texttt{fits}; source signals include pre-$T_0$ PRs \#18867 and
\#18818 and issue \#18815. &
Fix recurring \texttt{fits} correctness and edge-case failures. &
Strong: PR \#19438 fixed verification of \texttt{CompImageHDU} headers,
matching a FITS correctness family. \\
\texttt{python/mypy} &
Refactor anchor \texttt{librt}; source signals include pre-$T_0$ PRs \#20233,
\#20010, and \#19989. &
Refactor \texttt{librt} internals while preserving behavior. &
Strong: PR \#20623 made several \texttt{librt.internal} functions
positional-only, matching an internal cleanup direction. \\
\texttt{scrapy/scrapy} &
Refactor anchor \texttt{cleanup}; source signals include pre-$T_0$ PRs \#6802,
\#7043, and issue \#6924. &
Refactor \texttt{cleanup} internals while preserving behavior. &
Strong: PR \#7177 refactored the media-pipeline item handler to
\texttt{async def}, matching behavior-preserving internal modernization. \\
\bottomrule
\end{tabular}
\par\smallskip
\emph{Examples of forecast construction and retrospective validation.
Examples report validation evidence only; post-$T_0$ PR artifacts are not task
sources.}
\end{center}

\clearpage
\section{Task Construction Details}
\label{app:task-construction-details}

Task construction uses a multi-agent workflow so that task writing,
evaluation-asset construction, and executable verification remain distinct. The
role prompts below are simplified versions intended to show the construction
strategy for each role.

\paragraph{Context-selection role.}
The input is a validated forecast family, including its category, anchor,
expected behavior, target hints, and supporting pre-$T_0$ signal references,
together with the $T_{\mathrm{gen}}$ repository snapshot. The role searches the
repository using the anchor, spelling variants, target-module hints, and
category-specific words such as option, parser, cache, serializer, exception,
typing, or cleanup. It ranks candidate implementation files by anchor hits,
proximity to public APIs, import paths, and prior local test coverage, then
ranks candidate test files by whether they already exercise the same module,
public behavior, or CLI/API surface. It rejects rows whose only plausible targets
are documentation, packaging metadata, CI files, release scripts, or broad
project-wide rewrites.

\begin{center}
\begin{minipage}{0.93\linewidth}
\small
\hrule
\vspace{0.25em}
\textbf{Context-selection prompt skeleton.}
\begin{enumerate}[leftmargin=*,itemsep=0.08em,topsep=0.2em]
\item Inputs: forecast family, category, anchor, target hints,
$T_{\mathrm{gen}}$ file tree, and retrieved code/test excerpts.
\item Select 3--8 implementation candidates and 2--6 test candidates.
\item For each selected path, give a one-sentence repository-evidence rationale.
\item Choose one primary implementation target and one primary test target.
\item Reject instead of selecting when no focused code-and-test path exists.
\end{enumerate}
\vspace{0.1em}
\hrule
\end{minipage}
\end{center}

\paragraph{Task-writing role.}
The input is the forecast family and the selected $T_{\mathrm{gen}}$ repository
evidence. This role writes the model-visible issue-style request only. It turns
the family into a concrete request by naming the user-visible behavior, the
expected change, and the boundaries of the task. It must not see or mention
post-$T_0$ PR metadata, semantic-judge rationales, task-specific test-patch
content, gold-patch content, or execution logs. It avoids vague requests such as
``improve the module'' and broad rewrites such as ``redesign the parser''; the
request must be solvable from the repository snapshot and must include concrete
acceptance criteria.

\begin{center}
\begin{minipage}{0.93\linewidth}
\small
\hrule
\vspace{0.25em}
\textbf{Task-writing prompt skeleton.}
\begin{enumerate}[leftmargin=*,itemsep=0.08em,topsep=0.2em]
\item Inputs: forecast family, selected repository evidence, category, and
non-goals.
\item Write: title, repository, category, issue, requested change, acceptance
criteria, and forecast-conditioned context.
\item Make the request natural as a project issue and specific enough for a
coding agent.
\item Do not reveal test-patch logic, gold-patch logic, validation commands, or
post-$T_0$ PR evidence.
\item Reject if the request cannot be made focused, testable, and local.
\end{enumerate}
\vspace{0.1em}
\hrule
\end{minipage}
\end{center}

\paragraph{Oracle-design role.}
The input is the public task, selected implementation and test targets, and
repository test conventions. This role turns the public request into a concrete
oracle plan. It checks whether the behavior is already implemented in the
snapshot, identifies the smallest observable behavior that should fail before
the fix, chooses a test entry point, and specifies the expected failure and
pass-after condition. It rejects candidates when the intended behavior requires
external services, nondeterministic timing, large integration environments,
whole-repository migrations, or subjective maintainability judgments that cannot
be captured by a focused oracle.

\begin{center}
\begin{minipage}{0.93\linewidth}
\small
\hrule
\vspace{0.25em}
\textbf{Oracle-design prompt skeleton.}
\begin{enumerate}[leftmargin=*,itemsep=0.08em,topsep=0.2em]
\item Inputs: public task, selected files, relevant code excerpts, and test
style examples.
\item Define the target behavior in one sentence.
\item Specify the fail-before observation and pass-after observation.
\item Specify the test-patch location, test name, and setup assumptions.
\item Specify the smallest gold-patch scope, or reject with a reason.
\end{enumerate}
\vspace{0.1em}
\hrule
\end{minipage}
\end{center}

\paragraph{Patch-construction role.}
The input is the oracle plan and a clean $T_{\mathrm{gen}}$ checkout. The role
first writes the task-specific test patch, because the test patch defines the
target behavior. It then writes the gold patch, scoped to the smallest
implementation path that satisfies the oracle. The test patch may add a focused
regression test, structural check, or public-behavior check depending on the
task category. The gold patch must not include unrelated cleanup, formatting
churn, dependency updates, or historical PR content. If the task requires a
larger design change than the public request implies, the row is returned to
task writing or replaced.

\begin{center}
\begin{minipage}{0.93\linewidth}
\small
\hrule
\vspace{0.25em}
\textbf{Patch-construction prompt skeleton.}
\begin{enumerate}[leftmargin=*,itemsep=0.08em,topsep=0.2em]
\item Apply no historical PR patch and start from the clean
$T_{\mathrm{gen}}$ tree.
\item Write the task-specific test patch first; keep it focused on the oracle.
\item Run or reason about the expected fail-before behavior.
\item Write the gold patch with the smallest behavior-changing implementation.
\item Return both patches plus touched paths and any repair notes.
\end{enumerate}
\vspace{0.1em}
\hrule
\end{minipage}
\end{center}

\paragraph{Verification role.}
The input is the public task, task-specific test patch, gold patch, and a clean
checkout of the task-generation snapshot. Verification first checks that both
patches apply. It then applies only the test patch and expects the target oracle
to fail. Next, it applies the gold patch on top of the test patch and expects the
same oracle to pass. This is the \texttt{FAIL\_TO\_PASS} gate. Selected
\texttt{PASS\_TO\_PASS} checks are chosen from repository-native public tests
covering touched or nearby behavior. A row that fails target audit, leakage
review, patch application, \texttt{FAIL\_TO\_PASS}, or selected preservation
checks is returned for repair or replaced rather than admitted to satisfy the
200-task construction target.

\clearpage
\section{Synthesized Task Example}
\label{app:synthesized-task-example}

This appendix shows one complete synthesized task row. The example is a
bugfix task from \texttt{pallets/click}. It is conditioned on a validated
\texttt{click} bugfix family, but the public task, task-specific test patch, and
gold patch are constructed from the task-generation snapshot rather than from a
post-$T_0$ PR patch.

\begin{center}
\small
\begin{tabular}{@{}L{0.24\linewidth}L{0.66\linewidth}@{}}
\toprule
Field & Value \\
\midrule
Task id & \texttt{swe-future-v301-0001-click-choice-casefold-completion} \\
Repository & \texttt{pallets/click} \\
Category & Bugfix \\
Forecast family & Fix recurring \texttt{click} correctness and edge-case
failures. \\
Public behavior & Case-insensitive \texttt{Choice} conversion and shell
completion should use the same Unicode normalization. \\
Task-specific test & Add a focused shell-completion regression test for
\texttt{stra{\ss}e}/\texttt{STRAS}. \\
Gold patch & Replace \texttt{lower()} with \texttt{casefold()} in
\texttt{Choice.shell\_complete}. \\
\bottomrule
\end{tabular}
\end{center}

\paragraph{Model-visible public task.}
The following task text is visible to the evaluated coding agent.

\begin{lstlisting}
# Fix case-insensitive Choice completion for Unicode casefolding

Repository: `pallets/click`
Category: `bugfix`

## Issue

`click.Choice(..., case_sensitive=False)` uses Unicode `casefold()` when
converting command-line values, but its shell-completion path still compares
choices with `lower()`. This makes completion disagree with conversion for
valid non-ASCII choices such as `stra\u00dfe`, where `STRASSE` converts
successfully but the same prefix is not completed.

## Requested Change

Update `click.Choice.shell_complete` so case-insensitive completion uses the
same Unicode casefold normalization as `Choice.convert`. Keep the change local
to choice completion and preserve existing case-sensitive behavior.

## Acceptance Criteria

- `click.Choice(["stra\u00dfe"], case_sensitive=False).convert("STRASSE", None, None)`
  continues to return `"stra\u00dfe"`.
- The same choice returns `"stra\u00dfe"` from `shell_complete` for an incomplete
  value such as `"STRAS"`.
- Case-sensitive completions remain unchanged.
- Do not add broad rewrites, new dependencies, network calls, or unrelated
  cleanup.

## Forecast-Conditioned Context

This task instantiates the `click` bugfix forecast family by targeting a small
user-visible edge case in command-line completion behavior. The task is
generated from the repository snapshot and does not use any post-$T_0$ PR patch.
\end{lstlisting}

\paragraph{Task-specific test patch.}
The test patch defines the target oracle. With only this patch applied, the
task-generation snapshot should fail the new casefold-completion test.

\begin{lstlisting}
diff --git a/swe_future_tests/test_click_choice_casefold_completion.py b/swe_future_tests/test_click_choice_casefold_completion.py
new file mode 100644
--- /dev/null
+++ b/swe_future_tests/test_click_choice_casefold_completion.py
@@ -0,0 +1,22 @@
+from pathlib import Path
+import sys
+
+sys.path.insert(0, str(Path(__file__).resolve().parents[1] / "src"))
+
+import click
+
+def test_case_insensitive_choice_completion_uses_casefold():
+    choice = click.Choice(["stra\u00dfe"], case_sensitive=False)
+    assert choice.convert("STRASSE", None, None) == "stra\u00dfe"
+    completions = choice.shell_complete(None, None, "STRAS")
+    assert [item.value for item in completions] == ["stra\u00dfe"]
+
+def test_case_sensitive_choice_completion_is_unchanged():
+    choice = click.Choice(["Alpha"], case_sensitive=True)
+    assert [item.value for item in choice.shell_complete(None, None, "Al")] == ["Alpha"]
+    assert choice.shell_complete(None, None, "al") == []
\end{lstlisting}

\paragraph{Gold patch.}
The gold patch demonstrates that the task is solvable with a minimal local
implementation change.

\begin{lstlisting}
diff --git a/src/click/types.py b/src/click/types.py
--- a/src/click/types.py
+++ b/src/click/types.py
@@ -382,8 +382,8 @@ class Choice(ParamType, t.Generic[ParamTypeValue]):
         if self.case_sensitive:
             matched = (c for c in str_choices if c.startswith(incomplete))
         else:
-            incomplete = incomplete.lower()
-            matched = (c for c in str_choices if c.lower().startswith(incomplete))
+            incomplete = incomplete.casefold()
+            matched = (c for c in str_choices if c.casefold().startswith(incomplete))
 
         return [CompletionItem(c) for c in matched]
\end{lstlisting}

\paragraph{Executable validation.}
The task passes the executable release gates used by the constructor.

\begin{center}
\small
\begin{tabular}{@{}L{0.35\linewidth}L{0.22\linewidth}rL{0.18\linewidth}@{}}
\toprule
Gate & Expected result & Return code & Passed \\
\midrule
Public syntax check before patches & Pass & 0 & Yes \\
Apply task-specific test patch & Pass & 0 & Yes \\
Run target oracle before gold patch & Fail & 1 & Yes \\
Apply gold patch & Pass & 0 & Yes \\
Run target oracle after gold patch & Pass & 0 & Yes \\
Public syntax check after patches & Pass & 0 & Yes \\
Selected public preservation gate & Pass & n/a & Yes \\
\bottomrule
\end{tabular}
\end{center}

The selected public preservation gate for this row is a static public check. The
row records zero post-$T_0$ PR patch use. The evaluated model receives the
public task and repository snapshot; the task-specific test patch, gold patch,
and validation logs remain evaluation-side assets.

\bibliographystyle{plain}
\bibliography{references}

@inproceedings{jimenez2024swebench,
  title = {{SWE-bench}: Can Language Models Resolve Real-World GitHub Issues?},
  author = {Jimenez, Carlos E. and Yang, John and Wettig, Alexander and Yao, Shunyu and Pei, Kexin and Press, Ofir and Narasimhan, Karthik},
  booktitle = {International Conference on Learning Representations},
  year = {2024},
  url = {https://arxiv.org/abs/2310.06770},
  doi = {10.48550/arXiv.2310.06770}
}

@misc{wang2025swebenchpp,
  title = {{SWE-Bench++}: A Framework for the Scalable Generation of Software Engineering Benchmarks from Open-Source Repositories},
  author = {Wang, Lilin and Ramalho, Lucas and Celestino, Alan and Pham, Phuc Anthony and Liu, Yu and Sinha, Umang Kumar and Portillo, Andres and Osunwa, Onassis and Maduekwe, Gabriel},
  year = {2025},
  eprint = {2512.17419},
  archivePrefix = {arXiv},
  primaryClass = {cs.SE},
  url = {https://arxiv.org/abs/2512.17419},
  doi = {10.48550/arXiv.2512.17419}
}

@inproceedings{li2025feabench,
  title = {{FEA-Bench}: A Benchmark for Evaluating Repository-Level Code Generation for Feature Implementation},
  author = {Li, Wei and Zhang, Xin and Guo, Zhongxin and Mao, Shaoguang and Luo, Wen and Peng, Guangyue and Huang, Yangyu and Wang, Houfeng and Li, Scarlett},
  booktitle = {Proceedings of the Annual Meeting of the Association for Computational Linguistics},
  year = {2025},
  url = {https://arxiv.org/abs/2503.06680},
  doi = {10.48550/arXiv.2503.06680}
}

@misc{guo2026swefactory,
  title = {{SWE-Factory}: Your Automated Factory for Issue Resolution Training Data and Evaluation Benchmarks},
  author = {Guo, Lianghong and Wang, Yanlin and Li, Caihua and Tao, Wei and Yang, Pengyu and Chen, Jiachi and Song, Haoyu and Tang, Duyu and Zheng, Zibin},
  year = {2026},
  note = {To appear at FSE 2026},
  eprint = {2506.10954},
  archivePrefix = {arXiv},
  primaryClass = {cs.SE},
  url = {https://arxiv.org/abs/2506.10954},
  doi = {10.48550/arXiv.2506.10954}
}

@misc{zeng2026swehub,
  title = {{SWE-Hub}: A Unified Production System for Scalable, Executable Software Engineering Tasks},
  author = {Zeng, Yucheng and Li, Shupeng and Dong, Daxiang and Xu, Ruijie and Chen, Zimo and Zheng, Liwei and Li, Yuxuan and Zhou, Zhe and Zhao, Haotian and Tian, Lun and Xiao, Heng and Zhu, Tianshu and Hao, Longkun and Wu, Jianmin},
  year = {2026},
  eprint = {2603.00575},
  archivePrefix = {arXiv},
  primaryClass = {cs.AI},
  url = {https://arxiv.org/abs/2603.00575},
  doi = {10.48550/arXiv.2603.00575}
}

@misc{fu2026davincienv,
  title = {{daVinci-Env}: Open {SWE} Environment Synthesis at Scale},
  author = {Fu, Dayuan and Wu, Shenyu and Wu, Yunze and Peng, Zerui and Huang, Yaxing and Sun, Jie and Zeng, Ji and Jiang, Mohan and Zhang, Lin and Li, Yukun and Hu, Jiarui and Liu, Liming and Hou, Jinlong and Liu, Pengfei},
  year = {2026},
  eprint = {2603.13023},
  archivePrefix = {arXiv},
  primaryClass = {cs.SE},
  url = {https://arxiv.org/abs/2603.13023},
  doi = {10.48550/arXiv.2603.13023}
}

@misc{zhu2026sweplayground,
  title = {Training Versatile Coding Agents in Synthetic Environments},
  author = {Zhu, Yiqi and Gandhi, Apurva and Neubig, Graham},
  year = {2026},
  url = {https://github.com/neulab/SWE-Playground},
  note = {{SWE-Playground} project repository}
}

@misc{sun2026timeconsistent,
  title = {A Time-Consistent Benchmark for Repository-Level Software Engineering Evaluation},
  author = {Sun, Xianpeng and Sun, Haonan and Yu, Tian and Ma, Sheng and Zhang, Qincheng and Rao, Lifei and Tian, Chen},
  year = {2026},
  eprint = {2603.26137},
  archivePrefix = {arXiv},
  primaryClass = {cs.SE},
  url = {https://arxiv.org/abs/2603.26137},
  doi = {10.48550/arXiv.2603.26137}
}

@inproceedings{riddell2024contamination,
  title = {Quantifying Contamination in Evaluating Code Generation Capabilities of Language Models},
  author = {Riddell, Martin and Ni, Ansong and Cohan, Arman},
  booktitle = {Proceedings of the 62nd Annual Meeting of the Association for Computational Linguistics (Volume 1: Long Papers)},
  pages = {14116--14137},
  year = {2024},
  address = {Bangkok, Thailand},
  publisher = {Association for Computational Linguistics},
  url = {https://aclanthology.org/2024.acl-long.761/},
  doi = {10.18653/v1/2024.acl-long.761}
}

@inproceedings{matton2024leakage,
  title = {On Leakage of Code Generation Evaluation Datasets},
  author = {Matton, Alexandre and Sherborne, Tom and Aumiller, Dennis and Tommasone, Elena and Alizadeh, Milad and He, Jingyi and Ma, Raymond and Voisin, Maxime and Gilsenan-McMahon, Ellen and Gall{\'e}, Matthias},
  booktitle = {Findings of the Association for Computational Linguistics: EMNLP 2024},
  pages = {13215--13223},
  year = {2024},
  address = {Miami, Florida, USA},
  publisher = {Association for Computational Linguistics},
  url = {https://aclanthology.org/2024.findings-emnlp.772/},
  doi = {10.18653/v1/2024.findings-emnlp.772}
}

@misc{badertdinov2025swerebench,
  title = {{SWE-rebench}: An Automated Pipeline for Task Collection and Decontaminated Evaluation of Software Engineering Agents},
  author = {Badertdinov, Ibragim and Golubev, Alexander and Nekrashevich, Maksim and Shevtsov, Anton and Karasik, Simon and Andriushchenko, Andrei and Trofimova, Maria and Litvintseva, Daria and Yangel, Boris},
  year = {2025},
  eprint = {2505.20411},
  archivePrefix = {arXiv},
  primaryClass = {cs.SE},
  url = {https://arxiv.org/abs/2505.20411},
  doi = {10.48550/arXiv.2505.20411}
}

@article{graves2000faults,
  title = {Predicting Fault Incidence Using Software Change History},
  author = {Graves, Todd L. and Karr, Alan F. and Marron, J. S. and Siy, Harvey},
  journal = {IEEE Transactions on Software Engineering},
  volume = {26},
  number = {7},
  pages = {653--661},
  year = {2000},
  doi = {10.1109/32.859533}
}

@inproceedings{zimmermann2004versionhistories,
  title = {Mining Version Histories to Guide Software Changes},
  author = {Zimmermann, Thomas and Weissgerber, Peter and Diehl, Stephan and Zeller, Andreas},
  booktitle = {Proceedings of the 26th International Conference on Software Engineering},
  pages = {563--572},
  year = {2004},
  doi = {10.1109/ICSE.2004.1317478}
}

@inproceedings{nagappan2005churn,
  title = {Use of Relative Code Churn Measures to Predict System Defect Density},
  author = {Nagappan, Nachiappan and Ball, Thomas},
  booktitle = {Proceedings of the 27th International Conference on Software Engineering},
  pages = {284--292},
  year = {2005},
  doi = {10.1145/1062455.1062514}
}

@inproceedings{hassan2009changecomplexity,
  title = {Predicting Faults Using the Complexity of Code Changes},
  author = {Hassan, Ahmed E.},
  booktitle = {Proceedings of the 31st International Conference on Software Engineering},
  pages = {78--88},
  year = {2009},
  doi = {10.1109/ICSE.2009.5070510}
}

@article{kamei2013jit,
  title = {A Large-Scale Empirical Study of Just-in-Time Quality Assurance},
  author = {Kamei, Yasutaka and Shihab, Emad and Adams, Bram and Hassan, Ahmed E. and Mockus, Audris and Sinha, Anand and Ubayashi, Naoyasu},
  journal = {IEEE Transactions on Software Engineering},
  volume = {39},
  number = {6},
  pages = {757--773},
  year = {2013},
  doi = {10.1109/TSE.2012.70}
}

\end{document}